\documentclass[10pt,a4paper,twoside,fleqn]{article}
\usepackage{espcrc2}
\usepackage{robpreprintpage}
\usepackage[standard]{refstyle}

\def\ee{$e^+e^-$}               

\newcommand{\eref}[1]{(\ref{#1})}

\def\nbar{\bar n}            

\def\pt{p\kern -.2pt\lower 4pt\hbox{\tiny T}}    

\def\p0{P_0(\Delta y)}

\def\avg#1{\langle #1 \rangle}  

\def\NF{\mathcal{N}_{\kern -1.9pt f}}
\def\NC{\mathcal{N}_{\kern -1.7pt c}}

\usepackage[fleqn]{amsmath}
\usepackage{epsfig}

\DFTTtrue
\def\ksoft{k_{\text{soft}}}
\def\nsoft{\nbar_{\text{soft}}}
\def\ksemi{k_{\text{semi-hard}}}
\def\nsemi{\nbar_{\text{semi-hard}}}
\def\asoft{\alpha_{\text{soft}}}

\def\myitem#1{\medskip\noindent {#1}.~}

\title{Soft and semi-hard components structure in multiparticle
production in high energy collisions}
\author{A. Giovannini and R. Ugoccioni\address{
 Dipartimento di Fisica Teorica and I.N.F.N. - Sezione di Torino,\\ 
 via P. Giuria 1, 10125 Torino, Italy}\thanks{Work supported in part by
 M.U.R.S.T. under grant 1996}}

\begin{document}

\setabstract{Superposition 
of jets of different flavors as well as superposition of
jets of different topologies describe well observed structures (shoulder,
$H_q$ oscillations) in \ee\ annihilation. 
The analysis of similar effects   seen in experimental data in 
hadron-hadron collision is performed  successfully   by superimposing
soft  and  semi-hard  contributions at various energies.  
Negative Binomial
multiplicity distributions are chosen in all examined classes
of collisions as elementary substructures, i.e., as  QCD inspired genuine
self-similar fractal processes. 
Predictions on final particle multiplicity distributions in 
hadron-hadron 
collisions at 14 TeV are discussed.}

\thispagestyle{empty}

\ifDFTT
\makepreprinttitlepage{DFTT 65/97\\October 10, 1997}{
Soft and semi-hard components structure\\
in multiparticle production in high energy 
collisions}{A. Giovannini and R. Ugoccioni\\[0.3cm]
 \it Dipartimento di Fisica Teorica and I.N.F.N. - Sezione di Torino,\\ 
 \it via P. Giuria 1, 10125 Torino, Italy}{To be published in the
Proceedings of the\\XXVII International Symposium on Multiparticle
Dynamics\\Frascati (Italy), September 8--12, 1997\\[2cm] 
\normalsize Work supported in part by
 M.U.R.S.T. under grant 1996}
\fi

\maketitle
\ifDFTT\else \pagestyle{empty}\fi

\section{INTRODUCTION}

It is a fact, 1972 was   a quite important year for multiparticle dynamics.
It gave us

\myitem{1} QCD, the theory of strong interactions  \cite{QCD:QCD}, 
the non linear quantum 
field theory the past generation was looking for in order to explain 
particle production  firstly observed in cosmic ray physics in the 
thirties;

\myitem{2} KNO scaling behavior 
\cite{KNO}
for final particle    multiplicity distributions  (MD's) which can be 
considered an extension of  
previous results at  partonic level \cite{Polyakov};

\myitem{3} the first evidence that final particle multiplicity 
distributions in the 
accelerator region   
are broader than a Poisson distribution   in clear 
disagreement  with  multi-peripheral model   predictions
\cite{pp200};

\myitem{4} the negative binomial (NB) distribution --- 
called at the time Polya-Eggenberger   \cite{Polya} distribution ---
in order to take into account 
observed violations from multi-peripheralism  in hadron-hadron collisions.

\medskip
How important all these discoveries were and still are  in our field
is witnessed by the  fact that in our days, 25 years later,   in 
approaching the study of the elementary substructures in multiparticle 
production in the new foreseen energy domain in hadron hadron reactions 
(see FELIX project at CERN LHC) 
all above mentioned  results are to be used \cite{FelixLOI}.

To disentangle elementary substructures in the new horizon is indeed the 
only hope we have in order to reach  a simplified description of the 
expected complexity of final states and to provide accordingly a better
understanding of the subtle mechanisms controlling this extraordinary large
number of final particles production and related correlations.

Assuming our search successful, we have to answer  at least two subsequent
questions:
are the above mentioned elementary substructures common to all classes of
collisions?
and ---in addition--- are they really elementary?

Standard physical observables in this game are  
$n$ charged particle multiplicity 
distributions, $P_n$, 
factorial moments of the multiplicity distributions, $F_q$, 
and the  corresponding  cumulants moments, $K_q$, 
(these last observables are  
particularly sensitive to the tail  of the distribution where events with 
many particles give a relevant contribution). Of particular interest turns 
out to  be also the ratio $K_q/F_q$ known in the literature as  
$H_q$ variable.

All the mentioned  observables are not independent, 
but they are linked by the  following
equations  \cite{FrascatiRU,FaroRU}
\begin{gather}
  F_q = \sum_{n=q}^\infty n(n-1) \cdots (n-q+1) P_n
  \label{eq:Fq} \\ 
  K_q = F_q - \sum_{i=1}^{q-1} \binom{q-1}{i} K_{q-i} F_i
  \label{eq:Kq} 
\end{gather}

Eqs. \ref{eq:Fq} and \ref{eq:Kq} ---if interpreted correctly--- 
show   what is the main 
goal  of multiparticle  production, i.e., the integrated description of 
$n$-particle multiplicity distributions and of 
the corresponding correlations.
One should be able to relate the behavior of one observable to the germane 
behavior  of the others; in addition  the proposed explanation of one 
effect in one observable should also shed full light on the observed
effect in the others. 

Now, both in \ee\ annihilation  \cite{FrascatiRU} and in hadron-hadron 
collisions (as we shall see) two experimental  facts on $P_n$ and
$H_q$ observables in full phase space
attracted our attention:

\myitem{a} the  $P_n$ vs. $n$ peculiar behavior   
(shoulder structure of $n$-particle 
multiplicity distributions);

\myitem{b} the $H_q$ vs. $q$ oscillatory behavior 
($K_q/F_q$  $q$-particle correlations 
ratio oscillations).

\medskip
Here   it will be shown that both effects can be understood  in first
approximation in hadron-hadron collisions in terms of the same cause, i.e.,
as the weighted superposition of soft and semi-hard contributions.
This result  should be confronted  
with the explanation of  similar effects in   \ee\ annihilation 
by the weighted superposition   of  two- and multi-jets contributions. 

Two remarks. Although the superimposed  physical sub-structures in the two 
cases, \ee\ annihilation  and hadron-hadron collisions, are different
the weighted superposition mechanism is the same.

Secondly, all physical substructures are described by the same NB multiplicity
distributions and corresponding correlation functions, which are QCD inspired
genuine self-similar fractal processes.

\section{MULTIPLICITY DISTRIBUTIONS IN  HADRON-HADRON COLLISIONS  
IN THE GeV REGION}

In the accelerator region it has been shown that final $n$ charged particles 
multiplicity distribution in full phase space, $P_n$, is initially narrower 
than a Poisson distribution then it becomes Poissonian and finally 
at higher c.m.\ energies larger 
than a Poisson distribution. This behavior is usually described by using a 
two parameters distribution, the NBMD: 
the $k$ parameter  of the distribution 
is initially negative (the distribution is indeed a positive binomial) then 
it becomes infinite in correspondence of the the Poisson distribution 
and finally it is positive, i.e., it corresponds to a true NBMD.
Now, being $1/k$ closely related to the integral over full phase space of
the  two particle correlation function above considerations are favoring 
anti-correlations in the    lowest energy  domain ($k$ is negative, particles 
like  to stay far apart), independent particle production in the Poissonian 
regime and finally  two particle correlation dominance (hierarchical 
correlations structure) when the  multiplicity distribution is of NB type.

The problem is that  NB behavior  for final charged particles multiplicity
distributions can be trusted  in hadron-hadron collisions in full phase 
space only up to ISR energies. At higher  energies shoulder  structures  
start to  be clearly visible as shown by the UA5 Collaboration at CERN
$p\bar p$ collider.
The idea firstly suggested by C. Fuglesang  \cite{Fug} is  
to explain observed 
NB regularity violations as  the effect of the  weighted superposition of   
soft events (events without mini-jets)  and semi-hard events (events with 
mini-jets), the weight $\asoft$
being the fraction of soft events and the 
multiplicity distribution of each component being of NB type:
\begin{multline}
\hspace{-3mm} P_n(\asoft; \nsoft, \ksoft; \nsemi, \ksemi) = \\
	~~~~~\asoft P_n^{\text{NB}}(\nsoft, \ksoft) + \\
	(1 -\asoft) P_n^{\text{NB}}(\nsoft, \ksemi)	\label{eq:combo}
\end{multline}

\begin{figure}
\mbox{\epsfig{file=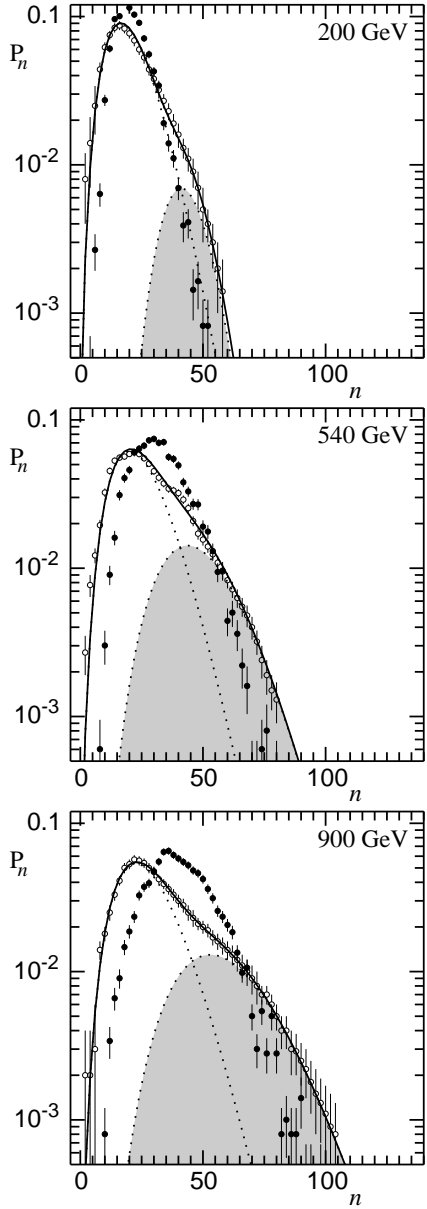,height=16cm}}
\caption[fig1]{Charged particle multiplicity distributions at c.m.\ energies
200 GeV, 546 GeV and 900 GeV. Data (white dots) are from UA5
Collaboration; the solid line is the fit with eq.~\protect\ref{eq:combo},
whose components are given by the dashed lines
(the semi-hard component is shaded). The solid dots show the
result of Monte Carlo calculations (Pythia).} \label{fig:Felix1}
\end{figure}

\begin{figure}
\mbox{\epsfig{file=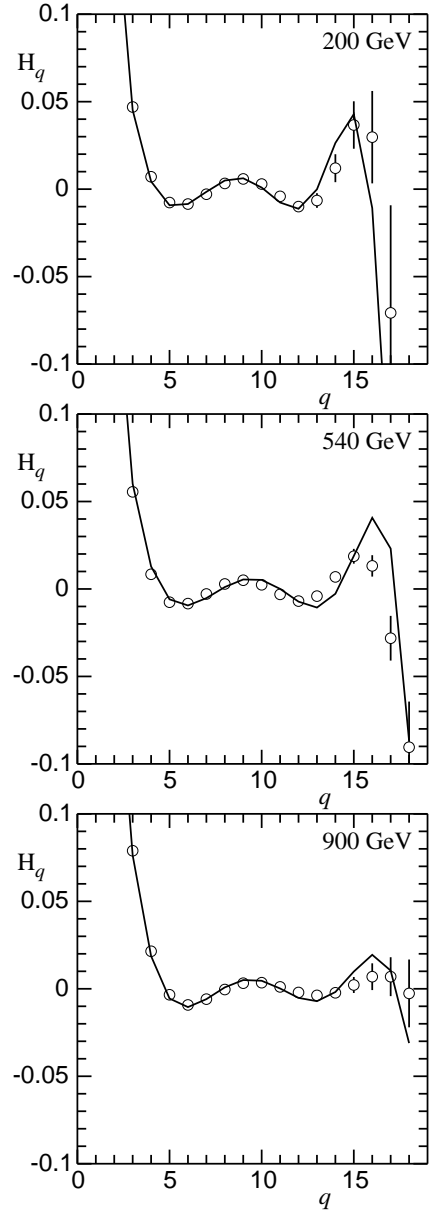,height=16cm}}
\caption[fig2]{$H_q$ vs $q$ at c.m.\ energies
200 GeV, 546 GeV and 900 GeV. Data (dots) are from the MD's in
Figure~\protect\ref{fig:Felix1}; the solid line is the prediction
of the fit also shown in Figure~\protect\ref{fig:Felix1}.

\vspace{3.05\baselineskip} 
} \label{fig:Felix2}
\end{figure}

In Fig. \ref{fig:Felix1} 
are shown the results of the compositions at 200 GeV, 560 GeV and
900 GeV c.m.\ energies  together with Pythia Monte Carlo calculations 
predictions; corresponding  parameters values of the two NB components are 
given in Table \ref{tab:Felix1}.
Notice that events with mini-jets (the shaded area in the figure) are
highly depressed with respect to events without mini-jets at 200 GeV
c.m.\ energy: mini-jets contribution grows quickly going to higher
c.m.\ energies.

\begin{table}
\caption{Values of the parameters of eq.~\protect\eref{eq:combo}
fitted to UA5 data \protect\cite{UA5:3}.
}\label{tab:Felix1}
\begin{center}
\begin{tabular}{llll}
\hline
c.m.\ energy & 200 GeV & 546 GeV & 900 GeV \\ 
\hline
$\asoft$ & 0.93 & 0.75 & 0.72 \\
$\nsoft$ & 19.9 & 24.0 & 26.9 \\
$\ksoft$ & 7    & 7    & 7    \\
$\nsemi$ & 42.2 & 47.6 & 57.9 \\
$\ksemi$ & 79   & 14   & 13   \\
\hline
\end{tabular}
\end{center}
\end{table}

\begin{table}
\caption[tab2]{NB observables used in the text.}\label{tab:Table2}
\def\somespace{\omit\vphantom{x}\cr}
\begin{center}
\begin{tabular}{l}
\hline
\somespace
$\displaystyle  P^{\text{NB}}_n = \frac{k(k+1)\dots(k+n-1)}{n!} 
	\frac{\nbar^n k^k}{(\nbar+k)^{n+k}} $ \cr
\somespace
$\displaystyle   F^{\text{NB}}_q = \nbar^q
   \frac{(k+1)(k+2)\dots(k+q-1)}{k^{q-1}}$ \cr
\somespace
$\displaystyle   K^{\text{NB}}_q = \nbar^q \frac{(q-1)!}{k^{q-1}}$\cr
\somespace
$\displaystyle   H^{\text{NB}}_q = \frac{(q-1)!}{(k+1)(k+2)\dots(k+q-1)}$ \cr
\somespace
\hline
\end{tabular}
\end{center}
\end{table}

The conclusion is that Pythia Monte Carlo calculations predictions on $P_n$
are unsatisfactory in the GeV region, whereas the 
proposed fit in terms of   the superposition of two NBMD's is quite good.
It is interesting to remark that  $H_q$ vs. $q$  oscillations 
(dots  in Figure \ref{fig:Felix2})
reconstructed from  the $n$ particle multiplicity distributions at different 
c.m.\ energies shown in Figure \ref{fig:Felix1} 
do oscillate and that the first minimum 
is  close to that observed in \ee\ annihilation. In addition by 
superimposing  weighted soft and semi-hard  contributions  $H_q$ vs. $q$ 
oscillations are quite  well described.

In conclusion the same cause explains both shoulder effects and $H_q$
oscillations as requested by an integrated description of multiplicity
distributions and correlations we emphasized in the introduction. 
What is really  striking  in our opinion is that elementary substructures 
of the weighted superposition mechanism in hadron - hadron collisions are 
again (as in \ee\ annihilation) NB  multiplicity distributions.
Analytic expressions of  $P_n$, $F_q$ and $K_q$ observables for the NBMD 
used in the text and in the figures are given in Table \ref{tab:Table2}.

\section{MULTIPLICITY DISTRIBUTIONS  AND $H_q$ OSCILLATIONS IN
HADRON-HADRON  COLLISIONS IN THE TeV REGION}

Above mentioned results are  important. The discovery that soft and 
semi-hard components can be  described  in terms of  NBMD's 
allows indeed to model 
on purely  phenomenological grounds and under very simple assumptions, 
expected scenarios for final particle multiplicity distributions 
and $H_q$ vs. 
$q$ oscillations in hadron hadron collisions in the new energy 
domain to be opened  at LHC. Accordingly  we  will study hadron-hadron 
collisions at  14 TeV c.m.\ energy, as expected by the LHC detector FELIX.

The point is to find  acceptable   energy dependence of the NB parameters  
$k$ and $\nbar$ for the two components substructures and the corresponding
weight factor $\asoft$. In a region where QCD has no predictions we shall 
proceed by phenomenological assumptions, which are 
discussed in the following.

The first assumption concerns energy dependence of the total average charged 
particle multiplicity, $\nbar$.
It is assumed here, as usually done \cite{giacomelli}
\begin{equation}
  \nbar(\sqrt{s}) = 3.01 - 0.474 \ln \sqrt{s} + 0.754 \ln^2 \sqrt{s}
					\label{eq:ntot}
\end{equation}

Since below 200 GeV c.m.\ energy one single NB fits multiplicity
distribution  
data very well and above 200 GeV c.m.\ energy soft component has been
disentangled  as shown in Table \ref{tab:Felix1},
  we propose to extrapolate the logarithmic 
increase with energy of the average charged particle multiplicity  of the
soft component,  $\nsoft$,   also at higher c.m.\ energy according to the 
formula
\begin{equation}
  \nsoft(\sqrt{s}) = -5.54 + 4.72 \ln \sqrt{s}		\label{eq:nsoft}
\end{equation}

For the average charged particle multiplicity of the semi-hard component,
$\nsemi$, we  then assume the  UA1 result on mini-jets  to be 
valid also at higher energy, i.e.,
\begin{equation}
  \nsemi(\sqrt{s}) \approx 2 \nsoft(\sqrt{s})		\label{eq:nsemi}
\end{equation}
It follows from
\begin{equation}
  \nbar= \asoft \nsoft + (1- \asoft)\nsemi         
\end{equation}
and Eqs.\ \ref{eq:nsoft} and \ref{eq:nsemi}, that
\begin{equation}
  \asoft = \frac{\nbar - \nsemi}{\nsoft - \nsemi} = 
	2 - \frac{\nbar}{\nsoft}
\end{equation}
In this way the weight factor $\asoft$ can be determined at 14 TeV and
it turns out to be equal to 0.30,
indicating a large mini-jets and hard jets production.

The second  problem concerns the energy dependence of the second NB
parameter, $k$.
In the accelerator region the parameter $k$ coincides with
$\ksoft$;  the
soft component is indeed the dominant one. Above 200 GeV the global 
distribution is not of NB type and parameter $k$ is hardly defined. However 
one can use the  dispersion $D^2 = \avg{n^2} - \avg{n}^2$ 
or equivalently the second factorial moment
\begin{equation}
  \frac{F_2}{\nbar^2} - 1 = \frac{D^2 - \nbar}{\nbar^2} = \frac{1}{k}
						\label{eq:F}
\end{equation}
                                                                         
Indications from our fit for the soft component at 200 GeV, 560 GeV and
900 GeV c.m.\ energy show that $\ksoft$ is constant above 200 GeV
c.m.\ energy.
In addition $1/\nsoft$ varies very little as the c.m.\ energy increases.
These two facts imply that KNO scaling should be valid for the soft 
component in a  very wide energy range; we have no reasons to believe that 
this behavior will change in the TeV region.
What about $\ksemi$?
Here two extreme   scenarios are possible.

\myitem{\textit{Scenario 1}} KNO scaling holds also for 
events of the semi-hard component, i.e.,
$D^2_{\text{semi-hard}} / \nsemi^2 = \text{const}$.
This fact with the remark that $1/\nsemi$ is a very small  quantity 
in the TeV region leads  to values of $1/\ksemi$ nearly constant 
but lower than $1/\ksoft$.
 Accordingly the resulting total $k$ value at 14 TeV is equal to 7.3 
($\ksemi$ is equal  to 13, whereas $\ksoft$ is equal to 7).
The ratio $D^2/\nbar^2$  after a quick increase from the accelerator to ISR 
region reaches its maximum at approximately one  TeV and then it decreases
towards its KNO limit  0.16 at 14 TeV.  

\myitem{\textit{Scenario 2}} Strong KNO scaling violation 
for events of the semi-hard
component, i.e., $D^2_{\text{semi-hard}} / \nsemi^2$
  is growing logarithmically
as already shown for the semi-hard component  in the  c.m.\ energy range 200 
GeV -- 900 GeV. We find $\ksemi$  to be equal to 3 and  $k$ total to 2.46  
at 14 TeV.   

\begin{figure*}
\mbox{\epsfig{file=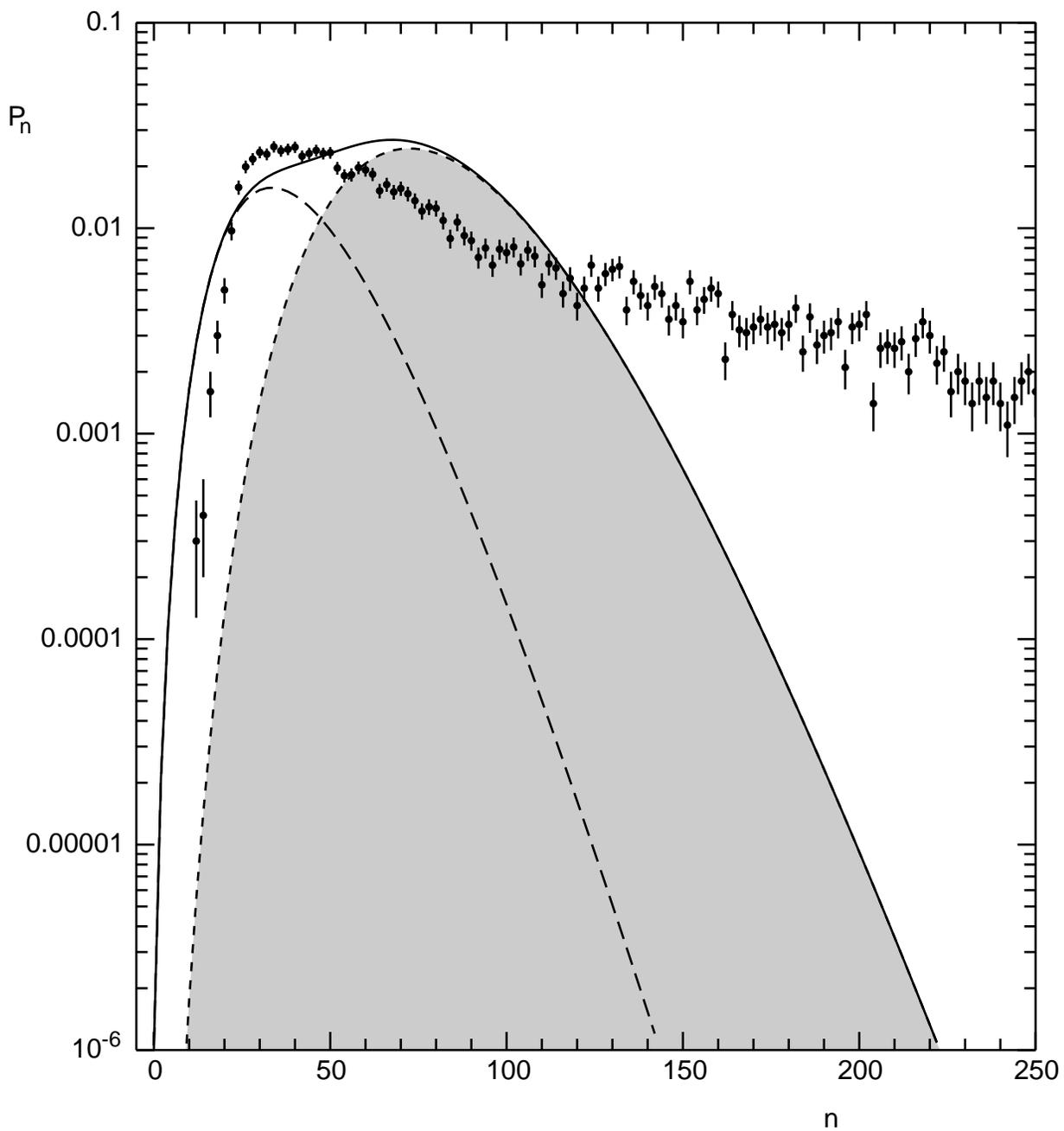,width=16cm}}
\caption{Charged particle multiplicity distributions at 14 TeV
for scenario 1 (KNO scaling dominance). The long- and short- dashed
lines represent the soft and the semi-hard component respectively
(the latter is also shaded),
the solid line represents the total MD. Dots are Pythia 
Monte Carlo predictions.
} \label{fig:Felix6}
\end{figure*}

\begin{figure*}
\mbox{\epsfig{file=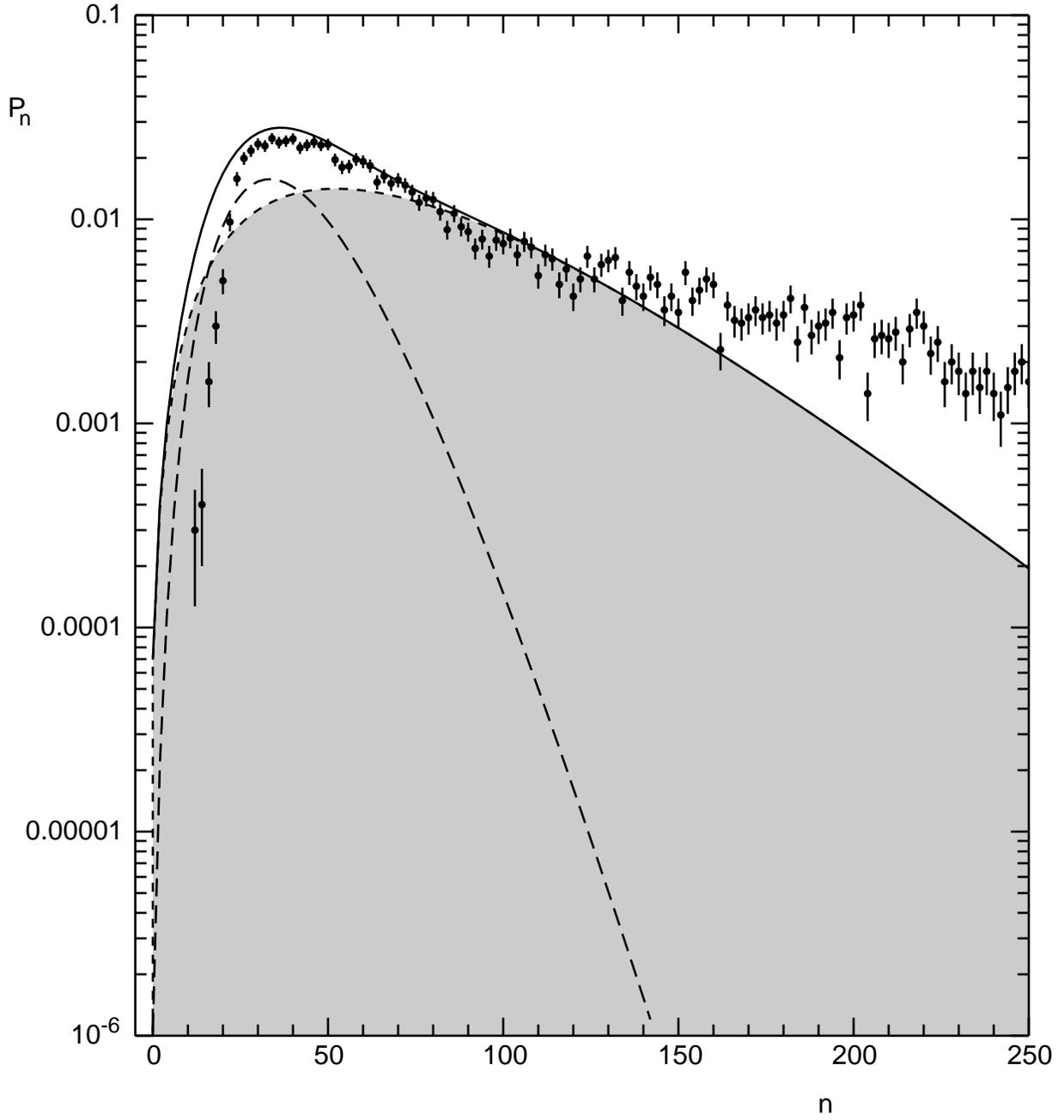,width=16cm}}
\caption{Same as Figure~\ref{fig:Felix6} but for scenario 2
(KNO scaling violation).
} \label{fig:Felix7}
\end{figure*}

\medskip
Resulting $n$ charged particles multiplicity distributions at 14 TeV for 
the  semi-hard  and soft components together  with the total multiplicity 
distribution for scenario 1 and 2 are shown in Figures 
\ref{fig:Felix6} and \ref{fig:Felix7} respectively.
Notice that the mini-jets fraction of events is much larger than the
soft events fraction.
Expectations from  Pythia Monte Carlo calculations at the same c.m.\ energy
are also given.
 The tail of the total multiplicity distribution is highly suppressed in 
scenario 1 when compared to MD of scenario 2, suggesting that independent 
particle production is favored  in the first  with respect to the second
scenario, where two particle correlations and larger mini-jets production
are dominant. Pythia Monte Carlo 
calculations  are closer to the soft component multiplicity distribution in 
scenario 1 in the low multiplicity region and to the semi-hard component 
multiplicity  distribution in the large multiplicity region in scenario 2.
Neither scenario reproduces Monte Carlo calculations for the total 
multiplicity  distribution. 

\begin{figure*}[t]
\mbox{\epsfig{file=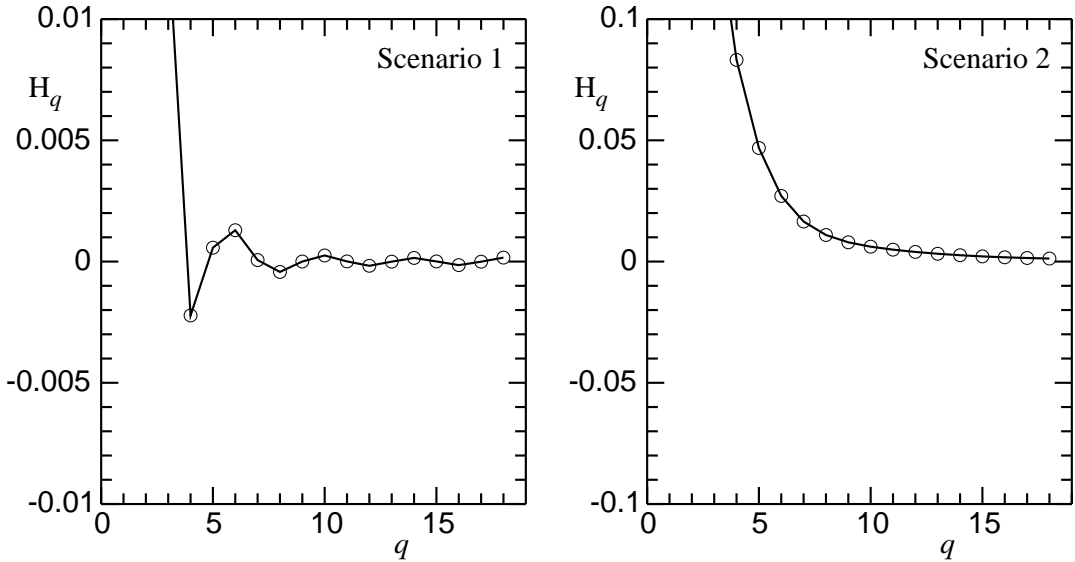,width=14cm}}
\caption{$H_q$ ratio vs $q$ for the scenarios described in the text.
The multiplicity distributions have not been truncated in order to
calculate the moments.}\label{fig:HqFrascati}
\end{figure*}

\begin{figure*}
\mbox{\epsfig{file=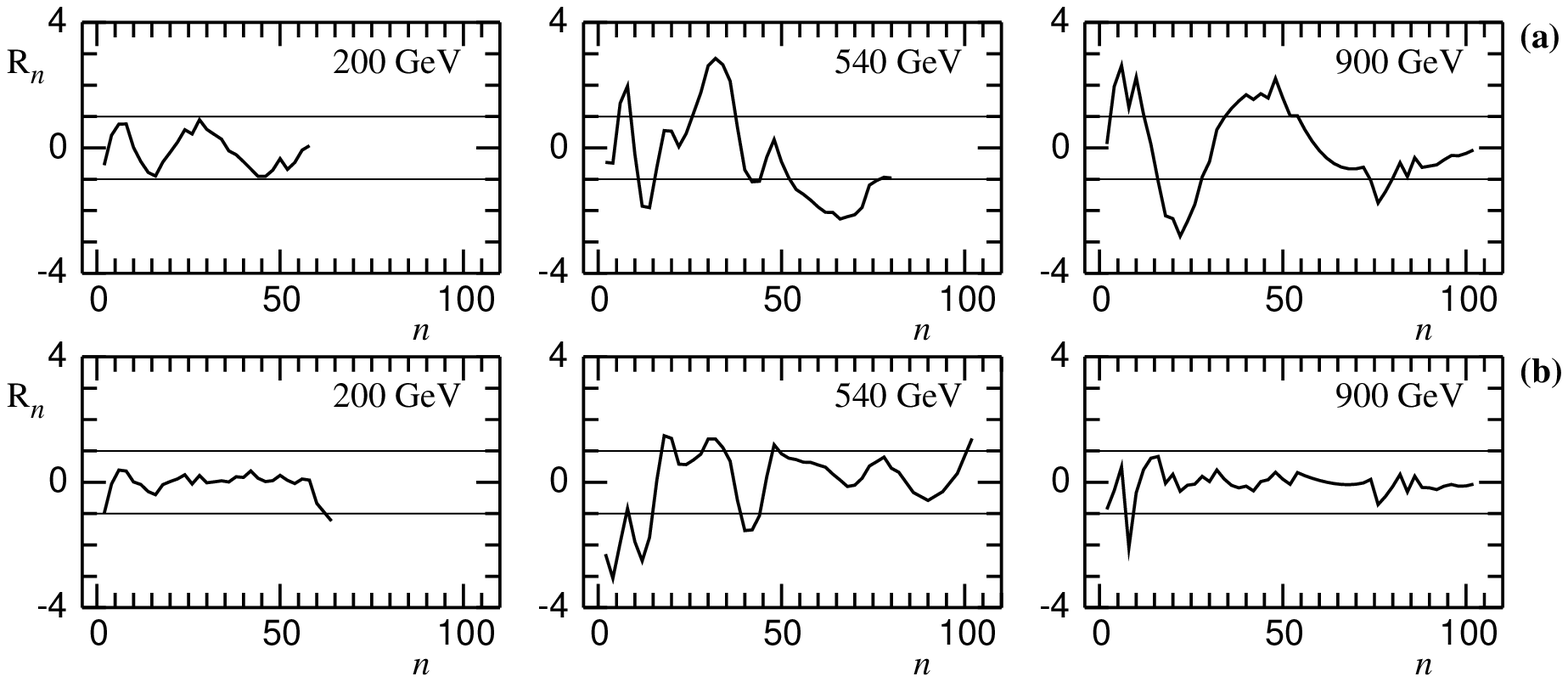,width=16cm}}
\caption[Fig6]{Residuals  for fits with (\textbf{a}) one single NBMD and 
(\textbf{b}) the weighted superposition of two NBMD's (shown in 
Figure \ref{fig:Felix1}) at c.m.\ energies of 200 GeV, 540 GeV
and 900 GeV.} \label{fig:Fig6}
\end{figure*}

In Figure \ref{fig:HqFrascati} 
$H_q$  vs. $q$ oscillations are also plotted in the two cases and
fitted by  $H_q$ values obtained  via  the weighed superposition of 
the two NBMD's.  It is interesting to remark that in the KNO scaling 
violation limit oscillations disappear.

One  question should still be answered: are the substructures which
we described in terms of  NBMD's really elementary?  
The lesson which we learned  from residuals analysis  of the  2-jet
sample of events in \ee\ annihilation which was indeed quite well
described by a single NB led us to the discovery
that this sample of events was  much better described by the weighted 
superposition of single jets of different flavor (each of them being
understood as a NBMD),  and to the prediction that single
jets are dominated by two particle correlations, which are flavor 
independent. It seems therefore quite natural in order to answer
to the above mentioned  question to proceed firstly to the residual 
analysis of hadron-hadron multiplicity distributions in full phase space 
in the highest c.m.\  energy region where data are available,  
i.e. again at 200 GeV, 560 GeV  and 900 GeV c.m.\ energies.

Results of this analysis  with one  NB fit to the experimental 
data are shown in  Figure 
\ref{fig:Fig6}a and with the weighted superposition of two
NBMD's (each of them corresponding to the soft and to the semi-hard component
respectively) in  Figure \ref{fig:Fig6}b. 
The improvement from Figure \ref{fig:Fig6}a to figure \ref{fig:Fig6}b is 
quite clear. This consideration notwithstanding we still see in the 
residuals of Figure \ref{fig:Fig6}b some additional substructures. 
When data  on 
multiplicity distributions in full phase space will be available at higher 
energies there  are good reasons to believe that the situation will be very
similar to that observed in the GeV region.

Then the initial  still unanswered question can be reformulated as follows:
is it possible to associate  the foreseen new elementary substructures in 
hadron-hadron  collisions to single jets and to describe them ---as done in 
\ee\ annihilation--- in terms of NBMD's?

Assuming the answer is yes, this fact would be really striking. It will
teach us that  a crucial step has been done in our understanding of
NB universality. After this discovery NB universality should be looked
for not in the full sample of events, but at the most elementary level of 
investigation ---as suggested also by our studies on \ee\
annihilation--- 
i.e., at single jet level in all classes of collisions
$\dots$ where one should 
find the domain of an effective 
self-similar Markov branching process ---as NBMD really is---  at work.

Assuming the answer is no, this fact would be even more interesting. It
will tell us that we reached in multiparticle  dynamics a completely
new  domain whose fundamental mechanisms controlling particle production
are all to be discovered.

\section{CONCLUSIONS}

$H_q$ vs $q$ oscillations and $P_n$ vs $n$ shoulder structure in 
hadron-hadron
collisions can be explained in terms of the same cause in the GeV  region,
i.e., the weighted superposition of elementary substructures which are
identified with soft events (events without mini-jets) and semi-hard
events (events with mini-jets) respectively. These results are then
extended to the TeV c.m.\ energy region in two extreme scenarios
characterized the first by KNO scaling behaviour and the second by
KNO scaling violation of the semi-hard component. Extrapolations from
both scenarios disagree with Pythia Monte Carlo calculation
predictions.
It is interesting to remark that $H_q$ vs.\ $q$ oscillations disappear
in the KNO scaling violation framework.
A large mini-jets and jets production is expected in both scenarios.
70 percent of events are indeed of this type: this number is one order
of magnitude larger than the corresponding number at 200 GeV c.m.\
energy.
In addition produced mini-jets in the KNO scaling violation framework
are expected to contain more particles than those in the KNO scaling
scenario.

These results are particularly interesting in view of the following
remarks.

\myitem{1} The same weighted superposition mechanism  of two components
explains both effects  
also in \ee\ annihilation.

\myitem{2} Above mentioned substructures (soft and 
semi-hard events contributions in 
hadron-hadron collisions and 2- and multi-jets  events contributions in 
\ee\ annihilation) are all well described in terms of NBMD's.

\myitem{3} Residuals analysis  of the 2-jet sample of 
events in \ee\ annihilation
reveals further substructures which we associate to single jets of different
flavor and describe  again in terms of NBMD's.

\myitem{4} Residuals analysis in hadron-hadron collisions  multiplicity
distributions  extrapolated in the TeV region reveals  further substructures
also in the  soft and semi-hard components, whose study is a challenging
problem for future hadron hadron detectors.

\medskip
The next step of our programme is to extend the present theoretical
investigation to rapidity intervals and impact parameter space.

\section*{AKNOWLEDGEMENTS}

We would like to express our thanks and gratitude to the INFN National
Laboratories organizing committee  and in particular to G. Capon and 
Giulia Pancheri
for the splendid work done for the success of the Symposium under  quite
difficult conditions. 
  

\end{document}